\documentclass[runningheads]{llncs}
\usepackage{graphicx}
\usepackage[]{algorithmicx}
\usepackage{algpseudocode}
\usepackage{algorithm}
\usepackage{float}
\usepackage{subfigure}
\usepackage{amsmath,amssymb}
\usepackage[utf8]{inputenc}
\usepackage{hyperref}
\usepackage{todonotes}

\newcommand{\onlysat}{\textup{\textsc{sat}}}
\newcommand{\sat}{\textup{\textsc{3-sat}}}
\newcommand{\maxsat}{\textup{\textsc{max-3-sat}}}
\newcommand{\qubo}{\textup{\textsc{qubo}}}
\newcommand{\choi}{\textup{\textsc{Choi}}$^{3m}$}
\newcommand{\chancellor}{\textup{\textsc{Chancellor}}$^{n+m}$}
\newcommand{\jonasone}{\textup{\textsc{N{\"u}{\ss}lein}}$^{2n+m}$}
\newcommand{\jonastwo}{\textup{\textsc{N{\"u}{\ss}lein}}$^{n+m}$}

\newtheorem{mydef}{Definition}

\begin{document}
\title{Solving (Max) 3-SAT via\\Quadratic Unconstrained Binary Optimization}
%
%\titlerunning{Abbreviated paper title}
% If the paper title is too long for the running head, you can set
% an abbreviated paper title here
%
\author{Jonas N\"{u}{\ss}lein \and  \and Thomas Gabor \and Sebastian Feld \and Claudia~Linnhoff-Popien}
%
% First names are abbreviated in the running head.
% If there are more than two authors, 'et al.' is used.
%

\author{Jonas Nüßlein\inst{1}\orcidID{0000-0001-7129-1237} \and
Sebastian Zielinski\inst{1} \and
Thomas Gabor\inst{1} \and
Claudia Linnhoff-Popien\inst{1} \and
Sebastian Feld\inst{2}
}
\authorrunning{N\"{u}{\ss}lein, Zielinski, Gabor et al.}
% First names are abbreviated in the running head.
% If there are more than two authors, 'et al.' is used.
%
\institute{Institute for Informatics, LMU Munich, Germany \\ \email{jonas.nuesslein@ifi.lmu.de} \and
Faculty of Electrical Engineering, Mathematics and Computer Science, TU Delft, Netherlands
}
\maketitle              % typeset the header of the contribution
\begin{abstract}
We introduce a novel approach to translate arbitrary {\sat} instances to Quadratic Unconstrained Binary Optimization ({\qubo}) as they are used by quantum annealing (QA) or the quantum approximate optimization algorithm (QAOA). Our approach requires fewer couplings and fewer physical qubits than the current state-of-the-art, which results in higher solution quality. We verified the practical applicability of the approach by testing it on a D-Wave quantum annealer.

\end{abstract}
\setlength{\parindent}{0pt}
\section{Introduction}

In recent years, many well-known optimization and decision problems have been translated to the model of quadratic unconstrained binary optimization (QUBO) \cite{lucas2014ising,glover2018tutorial}. The main motivation behind this is that QUBO models can be used as a problem specification for various early quantum algorithms, most notably the quantum approximate optimization algorithm (QAOA)~\cite{farhi2014quantum,zahedinejad2017combinatorial} and quantum annealing (QA)~\cite{kadowaki1998quantum,johnson2011quantum}. Current quantum computers are noisy and limited in size; thus it is important to encode problems as efficiently as possible. However, quantum hardware is conjectured to further grow in capability and a first demo application recently suggested that it might already have a substantial advantage over classical hardware for specific tasks~\cite{arute2019quantum}.

The most promising problems to be solved using quantum algorithms certainly include problems of the complexity class NP-hard, which are hard to solve for classical computers (unless $\textup{P} = \textup{NP}$)~\cite{cook2006p,fortnow2009status}. Many NP-hard problems like scheduling~\cite{venturelli2015quantum}, quadratic assignment~\cite{mcgeoch2013experimental}, or travelling salesman~\cite{feld2018hybrid} are of immense practical importance and practical instances often challenge current computing hardware. Thus, the eventual benefit of making these kinds of problems faster to solve may be especially appealing.

The canonical problem for the class NP-complete is 3-satisfiability (\sat), which we focus on in this paper~\cite{cook1971complexity}. A {\sat} instance is a formula in Boolean algebra and its solution is the binary answer to whether the formula is satisfiable.
\ \\

Our contributions in this paper are:

\begin{itemize}
    \item We present two novel {\sat}-to-{\qubo} translations: {\jonasone} and \newline {\jonastwo}
    \item We empirically show that {\jonasone} performs slightly better than {\chancellor} despite the bigger QUBO matrix
    \item We show that {\jonastwo} requires fewer couplings and fewer physical qubits than the current state-of-the-art approach {\chancellor}
    \item We empirically show that {\jonastwo} performs best, compared to three other {\sat}-to-{\qubo} translations
\end{itemize}

\section{Foundations}
\label{sec:foundations}

In this section, we introduce the mathematical foundations of the problems involved in the translation algorithms: {\sat} and {\qubo}.

\subsection{Satisfiability Problems}

The satisfiability problem ({\onlysat}) of propositional logic is informally defined as follows: Given a Boolean formula, is there any assignment of the involved variables so that the formula is reduced to ``true''?
The problem occurs in every application involving complex constraints or reasoning, like (software) product lines, the tracing of software dependencies, or formal methods~\cite{gabor2019assessing}.

All {\onlysat} problem instances can be reduced with only polynomial overhead to a specific type of {\onlysat} problem called {\sat}, in which the input propositional logic formula has to be in conjunctive normal form with all of the disjunctions containing exactly three literals.

\begin{mydef}[3-SAT]\label{def:3sat}
	A {\sat} instance with $n$ variables and $m$ clauses is given as (i)~a~list of variables $(v_j)_{0 \leq j \leq n-1}$, from which a list of literals $(l_i)_{0 \leq i \leq 3m-1}$ can be built of the form $$l_i \in \bigcup_{0 \leq j \leq n-1} \{v_j, \lnot v_j\},$$ and (ii) a list of clauses $(c_k)_{0 \leq k \leq m-1}$ of the form $$c_k = (l_{3k} \lor l_{3k+1} \lor l_{3k+2}).$$
	
	A given {\sat} instance is \emph{satisfiable} iff there exists a variable assignment given by the structure $(v_j \mapsto b_j)_{0 \leq j \leq n-1}$ with $b_j \in \{\top, \bot\}$ so that $$\bigwedge_{0 \leq k \leq m-1} c_k$$ reduces to $\top$ when interpreting all logical operators as is common. The problem of deciding whether a given {\sat} instance is satisfiable is called {\sat}.
\end{mydef}

%3-SAT is a variant of the satisfiability problem of propositional logic. A 3-SAT formula is a formula in conjunctive normal form, where each clause has a maximum of three literals. The task now is to find an assignment of the variables so that the formula is satisfied.
%\newline
%SAT can be polynomially reduced to 3-SAT. Thus 3-SAT is NP-complete after the Cook–Levin theorem.~\cite{cook}
%\newline

For example, we may write a {\sat} instance as Boolean formula $\mathcal{F} = ( a \lor b \lor c )$ $ \land $ $( a \lor \lnot c \lor \lnot d )$ consisting of $m=2$ clauses and featuring the $n=4$ distinct variables $\{a, b, c, d\}$. Obviously, $\mathcal{F}$ is satisfiable, for example via the variable assignment $(a \mapsto \bot, b \mapsto \top, c \mapsto \top, d \mapsto \bot)$.

{\sat} was the first problem to be shown to be NP-complete, which means that all problems in NP can be reduced to $\sat$~\cite{cook1971complexity}. In fact, as many proofs for NP-completeness for other problems build upon their reduction to {\sat}, {\sat} solvers can be used as tools to solve many different decision problems.

As {\sat} is central to many proofs of NP-completeness, it is somewhat surprising that when we generate random {\sat} instances with random amounts of variables $n$ and clauses $m$, most of these instances will be really easy to solve for standard {\onlysat} solvers. It is only as the ratio of clauses per variable approaches $\frac{m}{n} \approx 4.2$ that we can see the problems take exponential computing time. Knowing that many {\sat} instances are relatively easy to solve even for classical computers, we focus our attention regarding new methods (like quantum-based ones) on the critical {\sat} instances with $\frac{m}{n} \approx 4.2$.

{\maxsat} is an optimization problem that corresponds to the decision problem {\sat}. Instead of checking whether an assignment exists that fulfils the whole formula, i.e., reduces all clauses individually to $\top$, we try to find the assignment that fulfils as many clauses as possible. Note that {\maxsat} is a generalization of {\sat} as {\maxsat}'s optimal result is an assignment that fulfils all clauses and thus proves the satisfiability of the whole formula. 

\begin{mydef}[MAX-3-SAT]\label{def:max-3sat}
	A {\maxsat} instance is given the same way as a {\sat} instance (cf. Def.~\ref{def:3sat}). The objective of a {\maxsat} instance is to find a variable assignment of the structure $(v_j \mapsto b_j)_{0 \leq j \leq n-1}$ with $b_j \in \{\top, \bot\}$ so to $$\textit{maximize} \;\;\; \sum_{k=0}^{m-1} \;\; \begin{cases}1 &\textit{ if } c_k \textit{ reduces to } \top, \\ 0 &\textit{otherwise.}\end{cases}$$
\end{mydef}

%
%Example for a 3-SAT formula:
%\newline \newline
%$( a \lor b \lor c )$ $ \land $ $( a \lor \bar c \lor \bar d )$
%\newline
%Example-Solution:
%\newline \newline
%a = 0, b = 1, c = 1, d = 0
%\newline \newline
%
%Be in the following always:
%\begin{itemize}
%\item K = number of clauses of the 3-SAT formula
%\item V = number of variables of the 3-SAT formula
%\end{itemize}

\subsection{Quadratic Unconstrained Binary Optimization}

In quadratic unconstrained binary optimization (\qubo) we are looking for a binary vector $\textbf{x} = \langle x_i \rangle_{0 \leq i \leq k-1}$ of length $k$ that minimizes the value of a formula that at most contains quadratic terms in $x$.

\begin{mydef}[QUBO]
A {\qubo} instance with $k$ variables is given as a $k \times k$ matrix $Q \in \mathbb{R}^{k \times k}$. The objective of a {\qubo} instance is to find a binary vector $\mathbf{x} \in \mathbb{B}^k$ so to $$\textit{minimize} \;\;\; H(\mathbf{x}) = \sum_i Q_{ii}x_i + \sum_{i<j} Q_{ij}x_ix_j.$$
\end{mydef}

$H(\textbf{x})$ is also called the \emph{energy} of a {\qubo} solution $\textbf{x}$. Note that the lower triangle of the matrix $Q$ is always empty (since its values do not occur in the formula for the energy $H$). Finding the ideal solution vector $\textbf{x}$ of a given {\qubo} $Q$ is NP-hard.

When solving {\qubo} instances using a quantum annealer, the solution vector $\mathbf{x}$ is mapped to a set of qubits. These qubits have connections whose strength can be manipulated to emulate the values in the {\qubo} matrix. As the limiting factor in current hardware is the size of problems that can be solved, we seek translations to {\qubo} that require as few qubits (i.e., minimal size of the {\qubo} matrix) and as few connections between them (i.e., minimal density within the {\qubo} matrix) as possible.

\section{Related Work}
\label{sec:related}

There are currently two main approaches for translating {\sat} to {\qubo}, which we refer to as {\chancellor} \cite{chancellor2016direct} and {\choi} \cite{choi2010adiabatic}. We will review them in more detail in Subsections \ref{sec:chancellor} and \ref{sec:choi} respectively.
\ \\

In \cite{gabor2019assessing}, the authors examined the critical region of the problem domain for {\sat}, i.e., instances with $\frac{m}{n} \approx 4.2$. They observed that the clause-to-variable ratio has a great impact on the solution quality even on the quantum annealers.

Quantum annealing has previously been regarded as a solution to satisfiability problems: \cite{SATSolvingwithQA} focuses on embedding an originally {\onlysat}-related {\qubo} into the architecture of the most common quantum annealing chip. \cite{mooney2019mapping} shows a method to derive formulation for the optimization energy and proves mathematical bounds for the mapping of general $k$-\textsc{SAT} problems. Similarly, \cite{hen2016quantum} shows an approach justifying feasibility but provides no empirical data. In \cite{hogg2003adiabatic}, Grover's search algorithm was used to solve $k$-SAT. In \cite{nusslein2022algorithmic}, a {\qubo} formulation for $k$-SAT is proposed which only scales logarithmically in $k$ compared to the linear scaling in \cite{choi2010adiabatic} and \cite{chancellor2016direct}. In \cite{nusslein2022black}, a method is proposed to not hard-code a {\qubo} to {\onlysat} translation but to learn it using gradient-based methods.

\subsection{{\chancellor}}
\label{sec:chancellor}

Let $a_i^{(l)}$ be the $i$-th literal of clause $a^{(l)}$. The idea in \cite{chancellor2016direct} is to present a {\qubo} formulation for an arbitrary clause that assigns the energy $g$ to the \textit{one} variable assignment which does not fulfil the clause and the energy $0$ to all other possible variable assignments. The energy spectrum is therefore given by:

\[ \textit{Spec}(\{a^{(l)}\}) = \begin{cases} g & a_i^{(l)}=0 \;\; \forall i, \\ 0 &\textit{ otherwise.}\end{cases} \]

Thus we can create the {\qubo} formulation for the whole {\sat} formula by superimposing all clause-formulations: $H = \sum_l \textit{Spec}(\{a^{(l)}\})$. For $g>0$ the minimum energy bit-string will always be the one which satisfies the most clauses.
\ \\

To move from logical values to spin variables, one can map each logical variable $a_i=0$ to a spin variable with value $\sigma_i^z=-1$ and each logical variable $a_i=1$ to a spin variable with value $\sigma_i^z=+1$. Negation of the logical variable is then implemented through gauges on the spin variables. More precisely, $a_i$ is mapped to $c(i)\sigma_i^z$ with $c(i)=1$ and $\lnot a_i$ to $c(i)\sigma_i^z$ with $c(i)=-1$.
\ \\
 
The authors subsequently present the clause-formulation in the following way: The energy spectrum of the clause $(a_1 \lor a_2 \lor a_3)$ can be rewritten as $a_1 + a_2 + a_3 - a_1a_2 - a_1a_3 - a_2a_3 + a_1a_2a_3$ (with $a_i \in \{-1,+1\}$). The terms $a_1, a_2, a_3, a_1a_2, a_1a_3 $ and $a_2a_3$ can be directly inserted into the Ising Hamiltonian. For the triple term $a_1a_2a_3$, however, an ancilla qubit is necessary. The authors then present an Ising Hamiltonian for the triple term:

\[ H = h\sum_{i=1}^3 c(i) \sigma_i^z + J^a\sum_{i=1}^3 c(i) \sigma_i^z \sigma_a^z + h^a \sigma_a^z \]

in which up to the gauge choice $c(i) \in \{-1, 1\}$ the $3$ variables $\sigma_i^z$ are coupled with equal strength $J^a$ to the same ancilla spin variable $\sigma_a^z$.
\ \\

There are some constraints for the choice of the hyperparameters $h$, $J^a$, $h^a$, and $J$. We chose $h=g=1, h^a = 2h = 2,  J = 5$ and thus $J^a = 2J = 10$ as values for the variables in the ``specials cases'' section of Chancellor et al.~\cite{chancellor2016direct}. It is important to note that the choice of these values has no influence on the number of couplings needed. Any clause-translation will produce a fully-connected Ising/{\qubo} matrix (Note that Ising and {\qubo} are isomorphic).

For each clause exactly one ancilla qubit $C_i$ is needed. Thus the whole {\qubo} matrix will have size $n + m$. The {\sat} formula $( \lnot a \lor \lnot b \lor \lnot c ) \land ( a \lor b\lor c )$ would, for example, be represented by the {\qubo} matrix in \textbf{Table 1}.

\begin{table}[h]
\centering
\begin{tabular}[h]{|r||p{0.5cm}|p{0.5cm}|p{0.5cm}|c|c|}
\hline
& a & b & c & $C_1$ & $C_2$ \\
\hline
\hline
a & -88 & 48 & 48 & 40 & 40 \\
\hline
b & & -88 & 48 & 40 & 40 \\
\hline
c & & & -88 & 40 & 40 \\
\hline
$C_1$  & & & & -56 & 0 \\
\hline
$C_2$  & & & & & -64 \\
\hline
\end{tabular}
\vspace{1em}
\caption{{\qubo} matrix using {\chancellor} for the {\sat} formula $( \lnot a \lor \lnot b \lor \lnot c ) \land ( a \lor b\lor c )$.}
\end{table}

\subsection{{\choi}}
\label{sec:choi}

Choi \cite{choi2010adiabatic} provides a translation of {\sat} to {\qubo} that takes up $3m$ qubits, i.e., three qubits per clause in the original {\sat} formula (or one qubit per literal). It is inspired by the maximum independent set problem (to which {\sat} is first reduced, then to {\qubo}). Given a {\sat} instance with $m$ clauses and $n$ variables, {\choi} reserves a qubit $x_{k,i}, 0 \leq k < m, 0 \leq i \leq 2,$ for every literal. Thus {\choi} needs $3m$ qubits in total. One can interpret a solution candidate $x$ for this {\qubo} formulation in the following way:

\begin{itemize}
	\item If $x_{k,i} = 1$ and the corresponding literal $l_{3k+i} = v$ for some variable $v$, then we add the assignment $(v \mapsto \top)$ to the solution candidate for {\sat}.
	\item If $x_{k,i} = 1$ and the corresponding literal $l_{3k+i} = \lnot v$ for some variable $v$, then we add the assignment $(v \mapsto \bot)$ to the solution candidate for {\sat}.
	\item If $x_{k,i} = 0$, then we do nothing.
\end{itemize}

Note that a solution candidate for the {\qubo} may thus be illegitimate from the {\sat} perspective when it assigns different truth values to the same variable. Further note that {\choi}, even when returning the perfectly optimal solution, does not necessarily assign a truth value to every variable that occurs in the original formula.

For the detailed algorithm, we refer to~\cite{choi2010adiabatic} and will instead provide a small example. Note that the incentive and penalty values $X, Y, Z$ can be chosen rather freely as long as $Y > 2|X|$ and $Z > 2|X|$. Given the example {\sat} instance $(a \lor b \lor c) \land (a \lor b \lor \lnot c)$, we can then write a {\qubo} matrix as follows:

{\centering
$$(a \lor b \lor c) \land (a \lor b \lor \lnot c)$$
\begin{tabular}[h]{|r||c|c|c|c|c|c|}
\hline
$Q$ & $x_{0,0}$ & $x_{0,1}$ & $x_{0,2}$ & $x_{1,0}$ & $x_{1,1}$ & $x_{1,2}$ \\
\hline
\hline
$x_{0,0}$ & $-X$ & $Y$ & $Y$ & & & \\
\hline
$x_{0,1}$ & & $-X$ & $Y$ & & & \\
\hline
$x_{0,2}$ & & & $-X$ & & & $Z$ \\
\hline
$x_{1,0}$ & & & & $-X$ & $Y$ & $Y$ \\
\hline
$x_{1,1}$ & & & & & $-X$ & $Y$ \\
\hline
$x_{1,2}$ & & & & & & $-X$ \\
\hline
\end{tabular}

\hspace{2em}

}
\ \\

Intuitively, we need to penalize setting a pair of qubits from the same clause ($Y$) and penalize setting a pair of qubits which correspond to contradicting literals of the same variable ($Z$). Since so far we only assigned penalties, we need to set negative energy values on the diagonal ($-X$) in order to incentivize setting any qubits at all (and avoid the trivial solution $\textbf{x} = \mathbf{0}$).

\section{Approaches}
\label{sec:approaches}

We now describe two new approaches for translating a given {\sat} instance to {\qubo}. We introduce a new approach {\jonasone} in Section~\ref{sec:approaches:jonasone}, which uses $2n + m$ logical qubits, where $n$ is the number of variables and $m$ is the number of clauses. In Section ~\ref{sec:approaches:jonastwo} we then propose another formulation {\jonastwo}, which requires $n + m$ qubits. This is on par with the state-of-the-art {\chancellor}; however, {\jonastwo} uses fewer couplings, which leads to a reduction of physical qubits.

\subsection{A $2n + m$ Approach}
\label{sec:approaches:jonasone}

We now introduce a novel approach for the translation of {\sat} to {\qubo}: {\jonasone}. Like {\choi} and {\chancellor}, {\jonasone} actually solves {\maxsat} by trying to accumulate as many solvable clauses as possible. We build on the idea of \cite{nusslein2022algorithmic} to use an algorithm to describe the {\qubo} translation instead of an arithmetic notation.
\ \\

We use the qubits in the following way:
\begin{itemize}
	\item For each variable $v_j, 0 \leq j \leq n-1,$ occurring in the {\sat} instance, we use two qubits to encode if the variable is to be assigned $\top$ or if the variable is to be assigned $\bot$. Thus, $(v_j \mapsto \top)$ occurs in the variable assignment if $x_{2j} = 1$. Likewise, $(v_j \mapsto \bot)$ occurs in the variable assignment if $x_{2j+1} = 1$. Note that assigning both $v_{2j} = v_{2j+1}$ the same value makes for an illegitimate {\sat} solution candidate.
	\item Beyond those qubits, we further use one qubit for every clause in the {\sat} instance.
\end{itemize}

Effectively, the approach then uses $2n+m$ qubits for a {\sat} instance with $n$ variables and $m$ clauses. This may be less or more than the $3m$ qubits used in {\choi}; however, consider that difficult {\sat} instances are categorized by $\frac{m}{n} \approx 4.2$ (cf. Section~\ref{sec:foundations}). Thus, for the {\sat} instances which actually require extensive computations on classical computers, {\jonasone} manages to generate substantially smaller matrices. For the detailed instructions of Algorithm~\ref{alg:jonasone}, we first need to introduce the following definitions:

\begin{itemize}
\item We write $L = (v_0, \lnot v_0, ..., v_{n-1}, \lnot v_{n-1})$ for the list containing all possible literals given variables $(v_j)_{0 \leq j \leq n-1}$. Note that $|L| = 2n$.
\item We write $v_j \in c_k$ when clause $c_k$ contains a literal of the form $v_j$. Likewise, we write $\lnot v_j \in  c_k$ when $c_k$ contains a literal of the form $\lnot v_j$. We subsequently write $L_i \in c_k$ when $c_k$ contains the literal $L_i$.
\item We define $$R(L_{i}) = \sum_{k=0}^{m-1} \begin{cases} 1 & \textit{ if } L_i \in c_k, \\ 0 &\textit{ otherwise.}\end{cases}$$Thus  $R(L_i)$ is counting how often the literal $L_i$ occurs in the formula.
\item We define $$R(L_{i}, L_{i'}) = \sum_{k=0}^{m-1} \begin{cases} 1 & \textit{ if } L_i \in c_k \textit{ and } L_{i'} \in c_k,\\ 0 &\textit{ otherwise.}\end{cases}$$ Thus $R(L_{i}, L_{i'})$ is the number of occurrences of the literals $L_{i}$ and $L_{i'}$ together in the same clause.
\end{itemize}

\begin{algorithm}[tb]
\caption{\jonasone}\label{alg:jonasone}
\begin{algorithmic}[1]
\Procedure{\jonasone}{}
%\State VAR$[]$ = List of the literals $($x1,-x1,x2,-x2,x3,-x3,…$)$
%\State KLA$[][]$ = List of the clauses $($clause = list containing 3 literals$)$
\State $Q = \mathbf{0} \in \mathbb{R}^{2n +m \; \times \; 2n+m}$
\For{$i := 0 \textbf{ to } 2n+m$}
\For{$j := i \textbf{ to } 2n+m$}
\If {$i = j \textbf{ and } j < 2n$}
\State $Q_{ij} := -R(L_i)$
\ElsIf {$i = j \textbf{ and } j \geq 2n$}
\State $Q_{ij} := 2$
\ElsIf {$j < 2n \textbf{ and } j-i = 1 \textbf{ and } i \mod 2 = 0$}
\State $Q_{ij} := m+1$
\ElsIf {$i < 2n \textbf{ and } j < 2n$}
\State $Q_{ij} := R(L_i, L_j)$
\ElsIf {$j \geq 2n \textbf{ and } i < 2n \textbf{ and } l_i \textbf{ in } c_{j-2n}$}
\State $Q_{ij} = -1$
\EndIf
\EndFor
\EndFor
\State \Return $Q$
\EndProcedure
\end{algorithmic}
\end{algorithm}

Intuitively, {\jonasone} (cf. Algorithm~\ref{alg:jonasone}) encodes how many clauses are fulfilled by the solution. For example, if the minimal energy $H^*$ of a given {\jonasone}-{\qubo} is $-20$ this means that 20 clauses are fulfilled. If the formula, however, has more than 20 clauses this means that the formula is not satisfiable. We can consider the example formula $(a \lor b \lor \lnot c) \land (a \lor \lnot b \lor \lnot c)$ and its translation to {\qubo} using {\jonasone}:
\vspace{-1em}

{\centering
$$(a \lor b \lor \lnot c) \land (a \lor \lnot b \lor \lnot c)$$
\begin{tabular}[h]{|r||c|c|c|c|c|c|c|c|}
%\cline{2-9}
\hline
$Q$ & $\;a\;$ & $\lnot a$ & $\;b$\; & $\lnot b$ & $\;c\;$ & $\lnot c$ & \rotatebox{90}{$(a \lor b \lor \lnot c)\;\;$} & \rotatebox{90}{$(a \lor \lnot b \lor \lnot c)\;\;$} \\
\hline
\hline
$a$ & -2 & 3 & 1 & 1 & 0 & 2 & -1 & -1 \\
\hline
$\lnot a$ & & 0 & 0 & 0 & 0 & 0 & 0 & 0 \\
\hline
$b$ & & & -1 & 3 & 0 & 1 & -1 & 0 \\
\hline
$\lnot b$ & & & & -1 & 0 & 1 & 0 & -1 \\
\hline
$c$ & & & & & 0 & 3 & 0 & 0 \\
\hline
$\lnot c$ & & & & & & -2 & -1 & -1 \\
\hline
$\;\;(a \lor b \lor \lnot c)$ & & & & & & & 2 & 0 \\
\hline
$\;\;(a \lor \lnot b \lor \lnot c)$ & & & & & & & & 2 \\
\hline
\end{tabular}

\hspace{1.5em}

}

\subsection{An $n + m$ Approach}
\label{sec:approaches:jonastwo}

In this section we present {\jonastwo}, which is a {\sat} (again actually {\maxsat}) to {\qubo} translation, which only requires $n + m$ logical qubits. This is on par with {\chancellor}. However, we will show that our approach requires fewer couplings, which leads to a reduction of needed physical qubits in the hardware embedding.
\ \\

We use the qubits in the following way:
\begin{itemize}
	\item For each variable $v_j, 0 \leq j \leq n-1,$ occurring in the {\sat} instance, we use one qubit to encode the value it is assigned. Thus, $(v_j \mapsto \top)$ occurs in the variable assignment iff $x_j = 1$. This implies that $(v_j \mapsto \bot)$ occurs in the variable assignment iff $x_j = 0$.
	\item Beyond those qubits, we again use one qubit for every clause in the {\sat} instance.
\end{itemize}

For the algorithm, we start with an empty {\qubo} matrix as a canvas and then add specific patterns of values for each clause. As these pattern stack, we acquire the final value of $Q_{ij}$ as a sum of all stacked values. The algorithm thus needs to iterate over all clauses and repeatedly update the {\qubo} matrix while doing so. As we need to look at each clause individually, we can assume without loss of generality that all clauses are sorted, i.e., all negated literals appear as far towards the back of the clause as possible. This leaves us with only four possible patterns for clauses:
$$( a \lor b\lor c ), ( a \lor b \lor \lnot c ), ( a \lor \lnot b \lnot c ), ( \lnot a \lor \lnot b \lor \lnot c )$$

We now want to arrange the energy levels for each of the four cases such that a satisfied clause (no matter in which way it was satisfied, i.e., with one literal, with two, or with three) has the energy $H^*$ and the \textit{one} state which does not satisfy the clause has the energy $H^+ = H^* + 1$. See \textbf{Table 2} for all pattern matrices that might occur. The final {\qubo} matrix is then constructed by adding the pattern matrices' values to the cells in the {\qubo} matrix that correspond to the involved variables.
\begin{table}[t]
\centering
\subtable[$( a \lor b\lor c )$, $H^* = -1$]{%
\begin{tabular}[h]{|r||p{0.5cm}|p{0.5cm}|p{0.5cm}|c|}
\hline
& a & b & c & \rotatebox{90}{$( a \lor b ) \;\;$} \\
\hline
\hline
a & & 2 & & -2 \\
\hline
b & & & & -2 \\
\hline
c & & & -1 & 1 \\
\hline
$( a \lor b )$ & & & & 1 \\
\hline
\end{tabular}
}
\hspace{5em}
\subtable[$( a \lor b \lor \lnot c )$, $H^* = 0$]{%
\begin{tabular}[h]{|r||p{0.5cm}|p{0.5cm}|p{0.5cm}|c|}
\hline
& a & b & c & \rotatebox{90}{$( a \lor b ) \;\;$} \\
\hline
\hline
a & & 2 & & -2 \\
\hline
b & & & & -2 \\
\hline
c & & & 1 & -1 \\
\hline
$( a \lor b )$ & & & & 2 \\
\hline
\end{tabular}
}
\\
\subtable[$( a \lor \lnot b \lor \lnot c )$, $H^* = 0$]{%
\begin{tabular}[h]{|r||p{0.5cm}|p{0.5cm}|p{0.5cm}|c|}
\hline
& a & b & c & \rotatebox{90}{$( a \lor \lnot b ) \;\;$} \\
\hline
\hline
a & 2 & -2 & & -2 \\
\hline
b & & & & 2 \\
\hline
c & & & 1 & -1 \\
\hline
$( a \lor \neg b )$ & & & & \\
\hline
\end{tabular}
}
\hspace{5em}
\subtable[$( \lnot a \lor \lnot b \lor \lnot c )$, $H^* = -1$]{%
\begin{tabular}[h]{|r||p{0.5cm}|p{0.5cm}|p{0.5cm}|c|}
\hline
& a & b & c & \rotatebox{90}{$( \lnot a \land \lnot b \land \lnot c ) \;\;$} \\
\hline
\hline
a & -1 & 1 & 1 & 1 \\
\hline
b & & -1 & 1 & 1 \\
\hline
c & & & -1 & 1 \\
\hline
$( \neg a \land \neg b \land \neg c )$ & & & & -1 \\
\hline
\end{tabular}
}
\caption{Pattern matrices for the four different types of clauses.}
\label{tab:pattern-matrices}	
\vspace{-2em}
\end{table}
For a {\sat} formula with $p$ clauses where there are no negated literals and $q$ clauses where there are only negated literals, a variable assignment that satisfies the entire formula has the energy $H^* = - p - q$.

We can now consider the example formula $(a \lor b \lor c) \land (a \lor \lnot b \lor \lnot c)$ and its translation to {\qubo} using {\jonastwo}:

{\centering
$$(a \lor b \lor c) \land (a \lor \lnot b \lor \lnot c)$$
\begin{tabular}[h]{|r||c|c|c|c|c|}
\hline
$Q$ & $a$ & $b$ & $c$ & \rotatebox{90}{$(a \lor b)$} & \rotatebox{90}{$(a \lor \lnot b) \;\;$} \\
\hline
\hline
$a$ & $\;\;\; 0+2 \;\;\;$ & $\;\; 2-2 \;$ & $0+0$ & $-2$ & $-2$ \\
\hline
$b$ & & $0+0$ & $0+0$ & $-2$ & $2$ \\
\hline
$c$ & & & $-1+1$ & $1$ & $-1$ \\
\hline
$(a \lor b)$ & & & & $\;\;\;\;\; 1 \;\;\;\;\;$ & 0\\
\hline
$\;\; (a \lor \lnot b)$ & & & & & $\;\;\; 0 \;\;\;$ \\
\hline
\end{tabular}
\vspace{1.5em}

}

A possible optimal solution to this {\qubo} would be $\mathbf{x} = \langle 1, 0, 0, 1, 1 \rangle$, which corresponds to the variable assignment: $(a \mapsto \top, b \mapsto \bot, c \mapsto \bot)$ with the energy $H(\mathbf{x}) = -1$. Note that this {\qubo} matrix uses $5$ logical qubits and $6$ couplings (non-zero weights in the {\qubo} matrix). We can compare that to the {\chancellor} formulation, which requires $5$ logical qubits as well but $9$ couplings:

{\centering
$$(a \lor b \lor c) \land (a \lor \lnot b \lor \lnot c)$$
\begin{tabular}[h]{|r||c|c|c|c|c|}
\hline
$Q\;\;\;$ & $a$ & $b$ & $c$ & $C_1$ & $C_2$ \\
\hline
\hline
$a\;\;\;$ & $\;\;\; -88 \;\;\;$ & $\;\; 40 \;$ & $40$ & $40$ & $40$ \\
\hline
$b\;\;\;$ & & $-88$ & $48$ & $40$ & $40$ \\
\hline
$c\;\;\;$ & & & $-88$ & $40$ & $40$ \\
\hline
$\;\;\;C_1\;\;$ & & & & $\;\;\;\;\; -64 \;\;\;\;\;$ & 0\\
\hline
$\;\;\;C_2\;\;$ & & & & & $\;\;\; -64 \;\;\;$ \\
\hline
\end{tabular}
\vspace{1.5em}

}

Another notable feature of {\jonastwo} is the possibility to use the same clause-qubit for more than one clause. For example in the formula $(a \lor b \lor c) \land (a \lor b \lor \lnot c)$ the logical sub-formula $(a \lor b)$ appears in both clauses thus we just need one clause-qubit instead of two. In total, we thus would need $4$ logical qubits instead of $5$.

\section{Empirical Evaluation}

To empirically verify that {\jonastwo} requires fewer couplings than {\chancellor} we created random {\sat} formulas, applied both approaches, and counted the number of non-zero elements in the corresponding {\qubo} matrices. The results are shown in \textbf{Figure 1}. The x-axis describes the number of variables $V$ in the {\sat} formula. We then created random formulas with $\lceil 4.2V \rceil$ clauses. As can be seen in the charts, for both approaches the number of non-zero couplings in the {\qubo} matrices scales linearly in the number of variables $V$ of the {\sat} formula. However, {\jonastwo} only requires roughly $0.7$ of the couplings that {\chancellor} needs.

\begin{figure*}[h]
\centering
\minipage{0.84\textwidth}
  \centering
  \includegraphics[width=\linewidth]{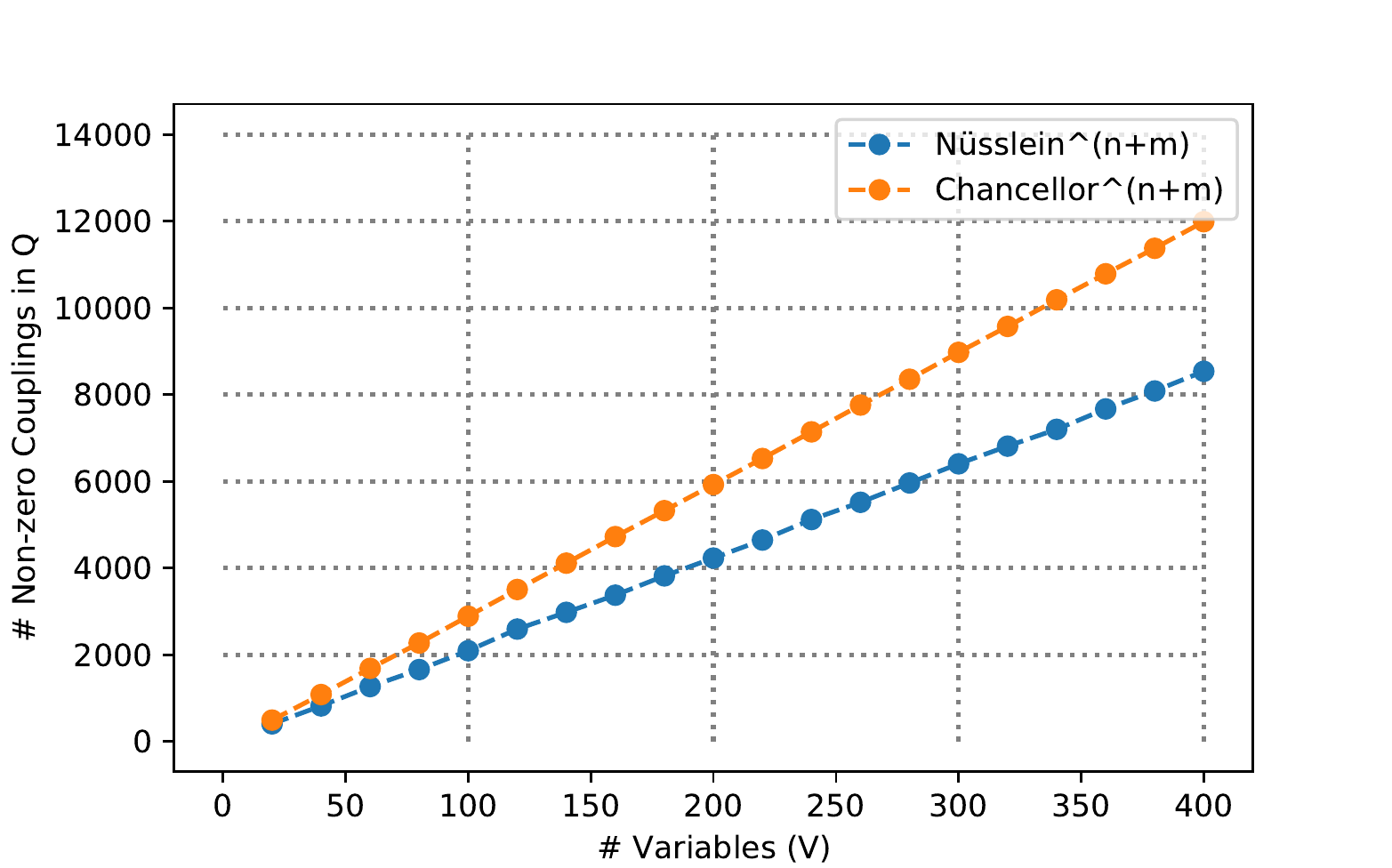}
  \caption{Relation of the number of variables in the {\sat} formula to the number of non-zero couplings in the {\qubo} matrix for the approaches {\chancellor} and {\jonastwo}.}\label{fig:awesome_image1}
\endminipage
\end{figure*}

In the next experiment, we evaluated how this reduction of couplings translates to a reduction of physical qubits. Note that both approaches {\jonastwo} and {\chancellor} require $n + m$ \textit{logical} qubits. However, to run a {\qubo} on a quantum annealer the {\qubo} has to be embedded into the hardware graph, which currently follows the Pegasus graph design \cite{boothby2020next}. We again created random {\sat} formulas for different $V$, applied both approaches to create the corresponding {\qubo} matrices and then ran the minorminer to find an embedding \cite{cai2014practical}. Finally, we counted how many \textit{physical} qubits were needed. The results (\textbf{Figure 2}) show that for both approaches the number of physical qubits scales linearly with $V$ but the chart of {\chancellor} has again a bigger gradient than the chart of {\jonastwo}. The line represents the median of 20 formulas and the shaded areas enclose the $0.25$ and $0.75$ quantiles.

\begin{figure*}[h]
\centering
\minipage{0.68\textwidth}
  \centering
  \includegraphics[width=\linewidth]{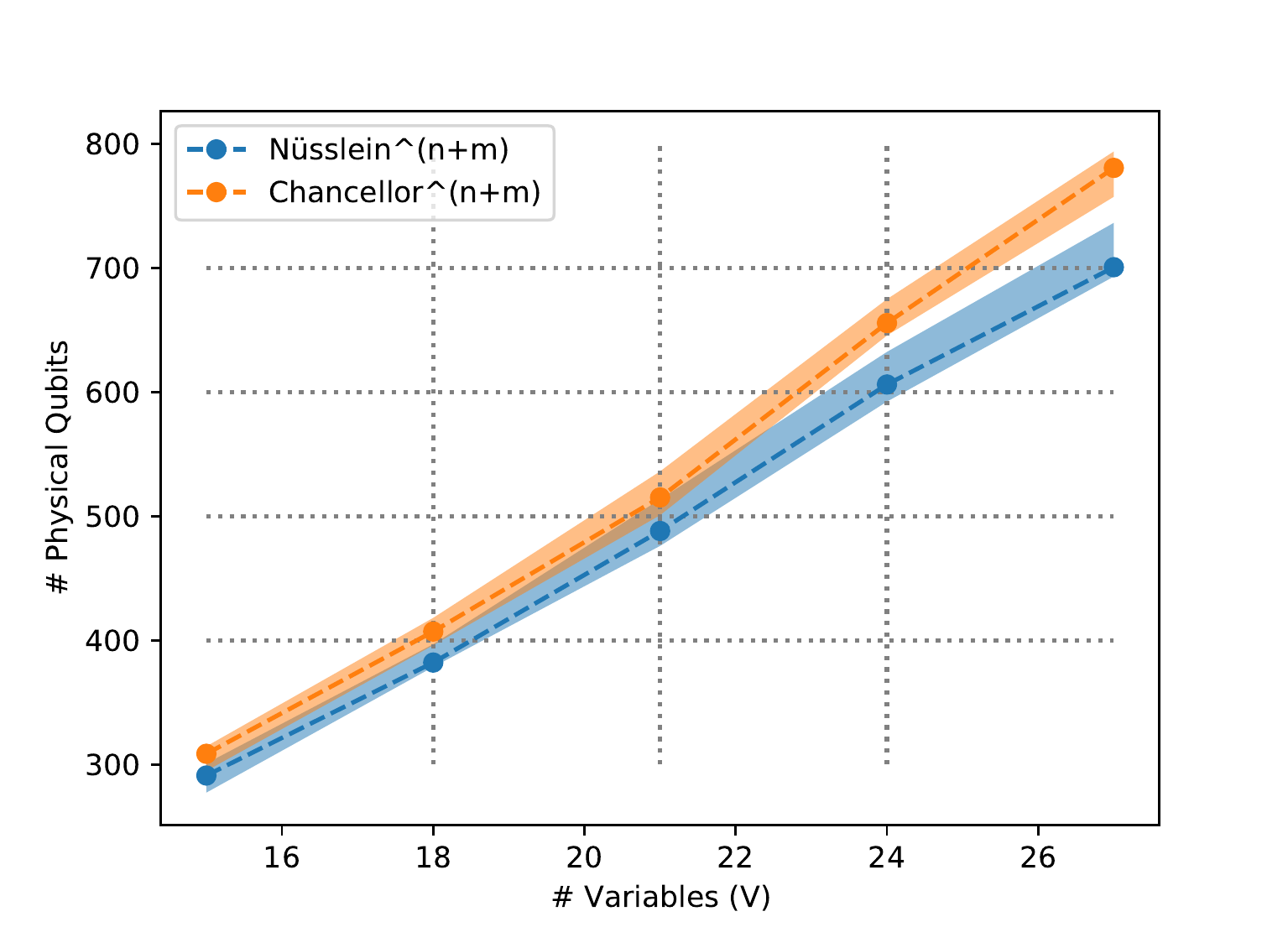}
  \caption{Relation of the number of variables in the {\sat} formula to the number of needed physical qubits for the approaches {\chancellor} and {\jonastwo}. The shaded areas enclose the $0.25$ and $0.75$ quantiles. }\label{fig:awesome_image1}
\endminipage
\end{figure*}

In a final experiment, we created random {\sat} formulas and solved them with all four methods on the D-Wave Quantum Annealer. We tested three sizes for the {\sat} formula and for each we created $20$ random formulas. \textbf{Table 3} shows the mean number of fulfilled clauses with the best-found variable assignment. For example for the size $(V=5, C=21)$ we created a random formula and solved it using {\jonasone} on the D-Wave. For the best answer of the D-Wave, we calculated the variable assignment and how many clauses are fulfilled with this assignment. We repeated this procedure for $20$ {\sat} formulas. As can be seen, {\jonastwo} was the best approach for every size of the formula. Another very interesting result is that {\choi} was mostly better than {\jonasone} and {\jonasone} was mostly better than {\chancellor} which indicates that the size of the {\qubo} matrix is not an optimal predictor for performance. The code for all four approaches can be found here: \href{https://github.com/JonasNuesslein/3SAT-with-QUBO}{https://github.com/JonasNuesslein/3SAT-with-QUBO}.

\begin{table}[h]
{\centering
\ \\
\begin{tabular}[h]{|r||c|c|c|}
\hline
 & $(V=5, C=21)$ & $(V=10, C=42)$ & $(V=12, C=50)$ \\
\hline
\hline
{\jonasone} & 20.4 & 39.0 & 45.0 \\
\hline
{\jonastwo} & \textbf{20.6} & \textbf{41.2} & \textbf{49.0} \\
\hline
{\chancellor} & 18.6 & 37.8 & 47.0 \\
\hline
{\choi} & 20.0 & 39.8 & 47.2 \\
\hline
\end{tabular}

\vspace{2em}

}
\caption{Performance of four {\sat} to {\qubo} translations on random formulas. The values represent the mean number of fulfilled clauses of the best-found solution vector.}
\label{table:1}
\end{table}

\section{Conclusion and Future Work}
\label{sec:conclusion}

In this paper, we presented two new approaches to translate {\sat} instances to {\qubo}. Despite the smaller size of the {\qubo}, the first approach {\jonasone} showed worse results than {\choi} in the experiments, which indicates that the size of a {\qubo} is not an optimal predictor for performance. For the other approach {\jonastwo}, we showed that it requires fewer couplings and fewer physical qubits than the current state-of-the-art {\chancellor}. We empirically verified that {\jonastwo} performs best compared to three other {\sat} to {\qubo} translations. The structure of the {\jonastwo} approach also shows a new paradigm in constructing {\qubo} translations: We did not derive a formulation from the original problem by adapting the mathematical framework; the {\qubo} matrix of {\jonastwo} was instead constructed from the ground up with the sole goal of mirroring {\sat}'s global optimum. We hope that {\jonastwo} can thus also inspire more new {\qubo} translations in the future.

Regarding {\sat}, it needs to be further investigated whether in general or for special cases even more favorable {\qubo} formulations for {\sat} exist. This investigation could also be formulated as an optimization problem (more precisely as Integer Linear Program), where all solutions of the {\sat} together with the energetically most favorable choice of auxiliary qubits must have the energy $H^*$ and all non-solutions together with the energetically most favourable choice of auxiliary qubits must have an energy $H^+ > H^*$. However, the choice of the ``most energetically favorable auxiliary qubits can be formulated by a series of linear inequalities (all other choices of auxiliary qubits with the same variable assignment must have a greater or equal energy). Following this approach, we may even be able to automatically generate new and efficient {\qubo} translations for practically relevant problems.
\ \\

%
% ---- Bibliography ----
%
% BibTeX users should specify bibliography style 'splncs04'.
% References will then be sorted and formatted in the correct style.
%
\bibliographystyle{splncs04}
\bibliography{references}
\end{document}